\documentstyle[preprint,aps,epsfig]{revtex}

\begin{document}
\draft 



\title {Second-harmonic generation in graded metallic films\footnote{This paper has appeard in Optics Letters {\bf 30}, 275 (2005).} }
\author{J. P. Huang$^{\#}$, K. W. Yu}

\address {Department of Physics, The Chinese University of Hong
Kong, Shatin, NT, Hong Kong}




\maketitle

\begin{abstract}

We study the effective second-harmonic generation (SHG)
susceptibility in graded metallic films by invoking the local
field effects exactly, and further numerically demonstrate  that the  graded metallic films can serve as a  novel optical material for producing  a broad structure in both the linear and SHG response and an
enhancement in the SHG signal. \\
{{\it ocis:} 160.4330, 310.6860,   160.4670, 160.4760.}
\end{abstract}




\newpage


There are much recent interest and practical need for nonlinear
optical materials that process large nonlinear susceptibility or
optimal figure of merit.  
  A large enhancement in
nonlinear responses has been found for a sub-wavelength multilayer
(i.e., thin film) of titanium dioxide and conjugated
polymer~\cite{Fischer}. 
For nonlinear effects other than the Kerr
effect,  Hui {\it et
al.}~\cite{Hui-JAP-1} derived general expressions for
the effective susceptibility for the second-harmonic generation
(SHG) in a binary composite of
random dielectrics.  They have also studied the thickness
dependence of effective SHG susceptibility in films of random
dielectrics and in composites with coated small particles~\cite{Hui04-1,Hui04-2}. Graded materials with various functionalities appear in nature and
in fabricated materials.  Graded thin films have many applications
as the gradation profile provides an additional control on the
physical properties. Graded thin films often possess different
optical properties~\cite{Grull}, when compared to
bulk materials. It is known that graded
materials have quite different physical
properties from the homogeneous materials~\cite{Milton}.
Also, it has been observed that the compositionally graded barium strontium titanate thin films have better electric properties than a single-layer barium strontium titanate film with the same compositions~\cite{LuAPL03}.
How to achieve  enhanced SHG is up to now a
challenging agenda~\cite{OL1,OL2}. 
In this paper, we shall investigate SHG in a graded metallic film
with an intrinsic SHG response and a graded linear response in the
metallic dielectric function. 



Let us consider a graded metallic film with thickness $L$, with
the gradation profile in the direction perpendicular to the film.
If we only include quadratic nonlinearities, the local
constitutive relation between the displacement field ${\bf D}(z)$
and the electric field ${\bf E}(z)$ in the static case would
be
$
D_i(z) = \sum_j \epsilon_{ij}(z)E_j(z) +
\sum_{jk}\chi_{ijk}(z)E_j(z)E_k(z) 
$~\cite{Hui04-1,Hui04-2}
with $i=x,y,z,$
where $D_i(z)$ and $E_i(z)$ are the $i$th component of  ${\bf
D}(z)$ and ${\bf E}(z)$, respectively, and $\chi_{ijk}$ is the
SHG susceptibility.
Here $\epsilon_{ij}(z)$ denotes the linear dielectric function,
which we assume for simplicity to be isotropic
$\epsilon_{ij}(z) = \epsilon (z)\delta_{ij}$.
Both $\epsilon(z)$ and $\chi_{ijk}(z)$ are functions of $z$ and
describe the gradation profiles.
In general, when a monochromatic external field is applied, the
nonlinearity will generate local potentials and fields at all
harmonic frequency. For a finite frequency external electric field
of the form
$
E_0 = E_{0}(\omega)e^{-i\omega t}+ {\rm c.c.},
$
the effective SHG susceptibility $\bar{\chi}_{2\omega}$ can be
extracted by considering the volume average of the displacement
field at the frequency $2\omega$ in the inhomogeneous
medium~\cite{Hui-JAP-1,Hui04-1,Hui04-2}. Next, we adopt a graded dielectric profile that
follows the Drude form
\begin{equation}
\epsilon(z,\omega) = 1 - \frac{\omega_p^2(z)}{\omega(\omega+i\gamma(z))},\label{graded} 
\end{equation}
where $0 \leq z \leq L$.  The general form in Eq.~(\ref{graded})
allows for the possibility of a gradation profile in the plasma
frequency [e.g., Eq.~(\ref{Wp})] and the relaxation rate [e.g., Eq.~(\ref{Rz})].  For a $z$-dependent profile, we can make
use of the equivalent capacitance for capacitors in series to
evaluate the effective perpendicular linear response for a given
frequency, 
$
1/\bar{\epsilon}(\omega) = (1/L)\int_0^L{\rm d}z [1/\epsilon(z,\omega)].
$
Using the continuity of the electrical displacement field, the
local electric fields $E(z,\omega)$ satisfies
\begin{equation}
\epsilon(z,\omega) E(z,\omega) = \bar{\epsilon}(\omega)
E_{0}(\omega),\label{relation}
\end{equation}
where $E_0(\omega)$ is the applied field along $z$ axis.
A $z$-dependent profile for the plasma frequency and the
relaxation time can be achieved experimentally. One possible way may
be to impose a temperature profile, as it has been observed that
surface enhanced Raman scattering is sensitive to
temperature~\cite{PRL94}.  Thus, one may tune the surface plasmon
frequency by imposing an appropriate temperature
gradient~\cite{Chiang1}.  A temperature gradient may also be used
in materials with a small band gap or with a profile on dopant
concentrations. In this case, one may impose a charge carrier
concentration profile to a certain extent.  This effect, when
coupled with materials with a significant intrinsic nonlinear
susceptibility, will give us with a way to control the effective
nonlinear response.  For less conducting materials, one may
replace the Drude form of dielectric constants by a Lorentz
oscillator form.  It may also be possible to fabricate dirty metal
films in which the degree of disorder varies in the $z$-direction
and hence leads to a relaxation-rate gradation profile.

The calculation of the effective nonlinear optical response then
proceed by applying the expressions derived in
Refs.~[2,3]. Next, the effective SHG susceptibility
$\bar{\chi}_{2\omega}$ is given by
$
\bar{\chi}_{2\omega} = \langle\chi_{2\omega}(z)
  E_{{\rm lin}}(z,2\omega)  E_{{\rm lin}}(z,\omega)^2 \rangle/[E_0(2\omega)E_0(\omega)^2], 
$
where $E_{{\rm lin}}$ is the linear local electric field in the
graded film with the same gradation profile but with a vanishing
nonlinear response at the frequency concerned. Using Eq.~(\ref{relation}) for
the linear local fields, the effective SHG susceptibility can be
rewritten as an integral over the film as
\begin{equation}
\bar{\chi}_{2\omega} = \frac{1}{L} \int_0^L {\rm d}z \chi_{2\omega}(z)
 \left(\frac{\bar{\epsilon}(2\omega)}{\epsilon(z,2\omega)}\right)
\left(\frac{\bar{\epsilon}(\omega)}{\epsilon(z,\omega)}\right)^2 .
\end{equation}


To illustrate the SHG in graded films, we consider as a model
system that the intrinsic SHG susceptibility $\chi_{2\omega}(z) =
\chi_{1}$ to be a real and positive frequency-independent constant
and does not have a gradation profile. In doing so, we are allowed to focus on
 the enhancement of the SHG response when compared to $\chi_1$. 
To show the effects of
gradation, here we take as a model plasma-frequency gradation profile
\begin{equation}
\omega_p(z)=\omega_p(0)(1-C_{\omega}\cdot z),\label{Wp}
\end{equation}
and a model relaxation-rate gradation profile~\cite{Neeves}
\begin{equation}
\gamma(z)=\gamma(\infty)+C_{\gamma}/z,\label{Rz}
\end{equation}
where $C_\omega$ and $C_{\gamma}$ are constant parameters tuning
the profile which is assumed to be linear in $z$.  Here
$\gamma(\infty)$ denotes the bulk damping coefficient, i.e., for
$z \rightarrow \infty$. Set thickness $L=1$ so that we could focus on  the film with a fixed thickness. Regarding the thickness dependence, we refer the reader to the work by Hui {\it et al.}~\cite{Hui04-1}

Figure~1 shows the real and imaginary parts of the effective
linear dielectric constant [Fig.~1(a) and Fig.~1(b)], and the real
and imaginary parts of the effective SHG susceptibility [Fig.~1(c)
and Fig.~1(d)] as a function of frequency $\omega/\omega_{p}(0)$.
Also shown is the modulus of $\bar{\chi}_{2\omega}/\chi_1$, see  Fig.~1(e). The
dielectric function gradation profile is given in Eqs.~(\ref{graded}),~(\ref{Wp})~and~(\ref{Rz}) with
$C_{\gamma}=0$, i.e., only a graded plasmon frequency is included. Throughout the calculations, the real part of the (linear)
dielectric constant is negative naturally.
 In this
case, a broad resonant plasmon band is observed. Note that for
$C_{\omega}\to 0$, $\omega_p(z)/\omega_p(0)\to 1$. 
As $C_{\omega}$
increases, $\omega_p(z)$ takes on values within a broader range
across the film, and leads to a broad plasmon band. Increasing
$C_{\omega}$ also causes the plasmon peak to shift to lower
frequencies. The reason is that, in analogous to capacitors in
series, the effective dielectric constant of the film is dominated
by the layer with the smallest dielectric constant. 
For the SHG response, the
frequency dependence is quite complicated.  As $C_{\omega}$
increases, it is noted that structures in the SHG response also
shifts to the lower frequencies and the range of values of the SHG
susceptibility increases as well.
Figure~2 displays the results of model calculations in which a
gradation profile of the relaxation rate of the form in Eq.~(\ref{Rz}) is
also included. The effects are similar to those in Fig.~1. The SHG
response is found to be enhanced more strongly in the presence of
both a relaxation-rate gradation and a plasma-frequency gradation
(see Fig.~1) than for plasmon-frequency gradation alone,
especially at low frequencies. As $C_{\gamma}$ increases, the
structures in the linear and SHG response both show a shift to
lower frequencies.
In Figs.~1-2, the quantities which can be both positive and negative are plotted in logarithm of modulus. When the quantities pass through zero, the logarithm will be very large, thus yielding spikes. In addition, we have used the normalized numbers in order to describe a general origin of the SHG in metal films rather than a specified metal film.


 The point for achieving the present results is that one needs a sufficiently
large gradient rather than a crucially particular form of the dielectric function or gradation profiles.    Thus, it is expected that an enhancement in SHG responses will
also be found in compositionally graded metal-dielectric composite
films in which the fraction of metal component varies
perpendicular to the film. 
 In the present work, due to the symmetry of the film, we have only enhancement for the
polarization perpendicular to the film (i.e., parallel to the direction of the gradient). In this polarization, the tangential
component of electric field $E$ vanishes identically. Thus, the continuity of normal
component of  $D$ [see Eq.~(\ref{relation})] gives rise to the enhanced SHG. However, for the polarization
parallel to the film, i.e., the tangential component does not vanish. In
this polarization, it is the continuity of the tangential component of $E$
that leads to no enhancement at all~\cite{Fischer}.


{\it Acknowledgments}.
We thank Professor Pak-Ming Hui
for fruitful discussions and suggestions, as well as a critical reading of
the manuscript. This work was
supported by the Research Grants Council of the Hong Kong SAR
Government. J. P. Huang's e-mail address is jphuang@mpip-mainz.mpg.de.

 $^{\#}$Present address: Max Planck Institute for
Polymer Research, Mainz, Germany.

\newpage

\begin{figure}[h]
\caption{(a) ${\rm Re}[\bar{\epsilon}(\omega)]$,   (b) ${\rm
Im}[\bar{\epsilon}(\omega)]$ (linear optical absorption), (c)
${\rm Re}[\bar{\chi}_{2\omega}/\chi_1]$,  (d) ${\rm
Im}[\bar{\chi}_{2\omega}/\chi_1]$, and (e) Modulus of
$\bar{\chi}_{2\omega}/\chi_1$  
versus the normalized incident angular frequency
$\omega/\omega_p(0)$ for the dielectric function gradation profile
 [Eq.~(\ref{graded})]
with various plasma-frequency gradation profile
 [Eq.~(\ref{Wp})]  and
relaxation-rate gradation profile
  [Eq.~(\ref{Rz})]. Here
$|\cdots |$ denotes the absolute value or modulus of $\cdots$.
Parameters: $\gamma(\infty)=0.02 \omega_p(0)$ and $C_\gamma=0.0$. }
\end{figure}

\begin{figure}[h]
\caption{Same as Fig.1. Parameters: $\gamma(\infty)=0.02
\omega_p(0)$ and $C_{\omega}=0.6$. }
\end{figure}

\newpage

\centerline{\includegraphics[width=6.0cm]{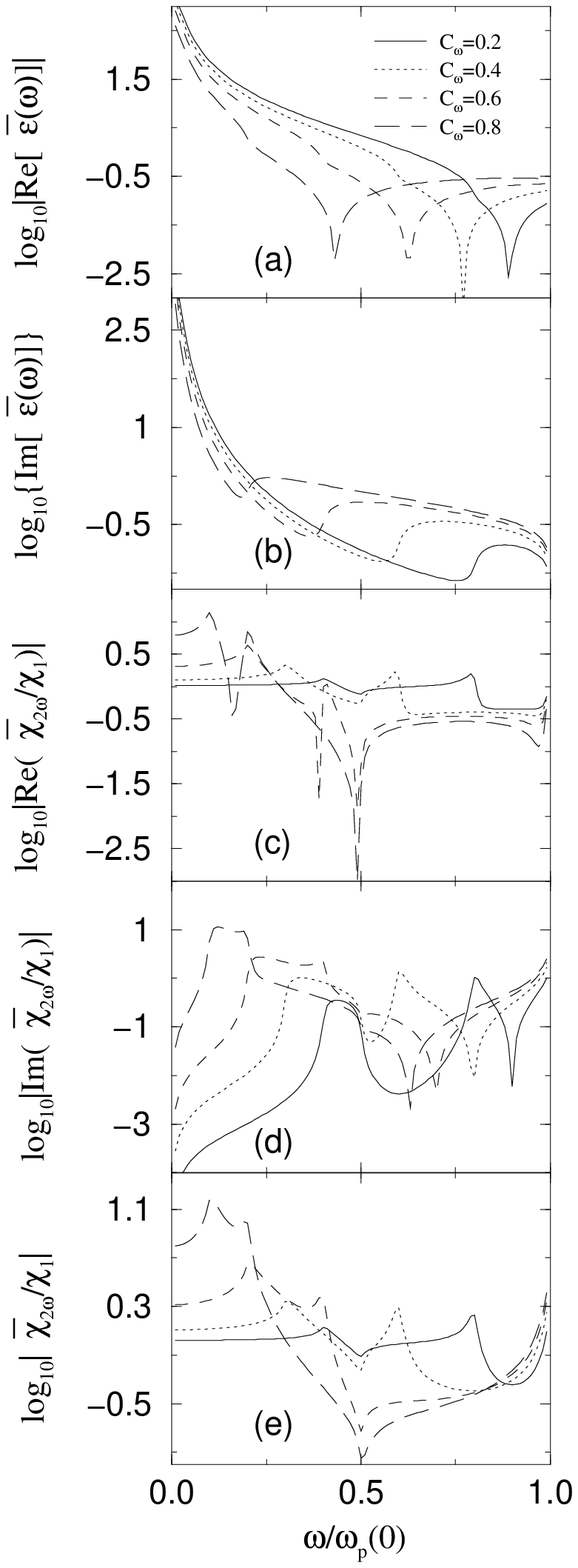}}
\centerline{Fig.~1/Huang and Yu}

\newpage

\centerline{\includegraphics[width=6.0cm]{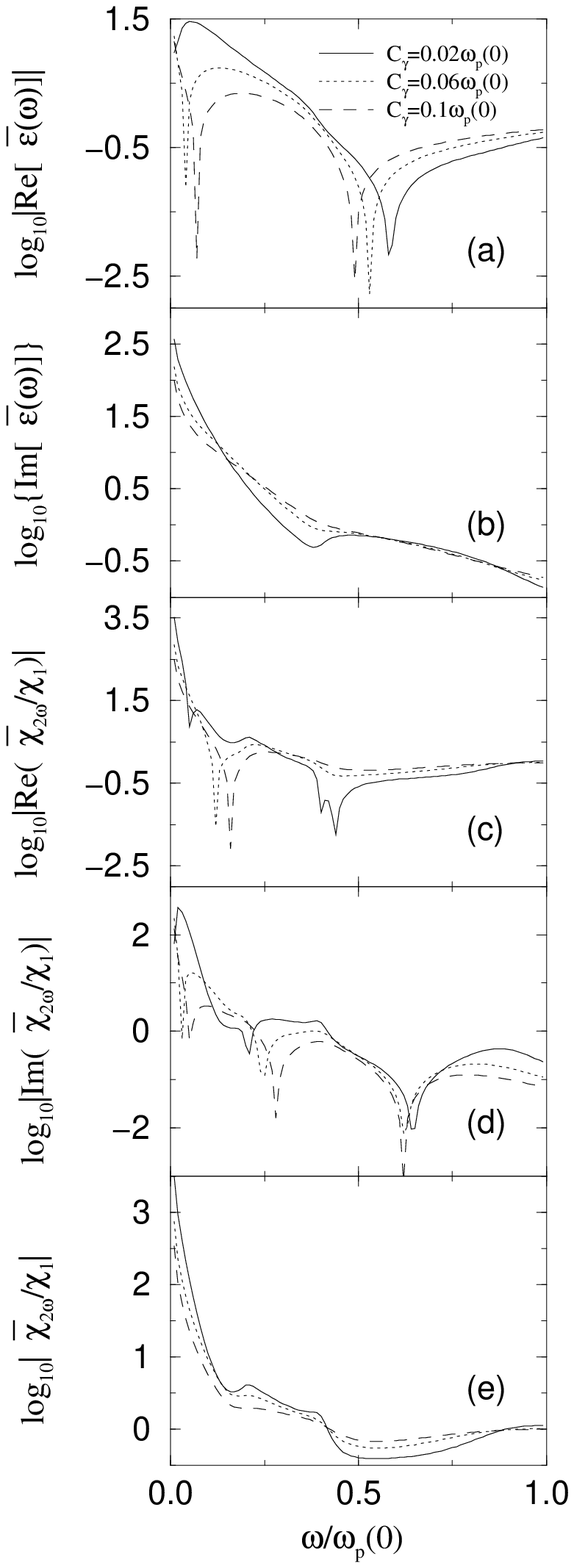}}
\centerline{Fig.~2/Huang and Yu}


\begin{references}


\bibitem{Fischer} G.L. Fischer, R.W. Boyd, R.J. Gehr, S.A. Jenekhe,
 J.A. Osaheni, J.E. Sipe, and L.A. Weller-Brophy,
 Phys. Rev. Lett. {\bf 74}, 1871 (1995).

\bibitem{Hui-JAP-1} P.M. Hui and D. Stroud, J. Appl. Phys. {\bf 82}, 4740 (1997).

\bibitem{Hui04-1} P.M. Hui, C. Xu, and D. Stroud, Phys. Rev. B {\bf 69}, 014202 (2004).

\bibitem{Hui04-2}  P.M. Hui, C. Xu, and D. Stroud, Phys. Rev. B {\bf 69}, 014203 (2004).

\bibitem{Grull} H. Gr\"{u}ll, A. Schreyer, N.F. Berk, C.F. Majkrzak, and C.C. Han, Europhys. Lett. {\bf 50}, 107 (2000).

\bibitem{Milton} G.W. Milton, {\it The Theory of Composites} (Cambrige University Press, Cambridge, 2002).



\bibitem{LuAPL03} S.G. Lu, X.H. Zhu, C.L. Mak, K.H. Wong,
 H.L.W. Chan, and C.L. Choy, Appl. Phys. Lett. {\bf 82}, 2877 (2003).

\bibitem{OL1} D. Pezzetta, C. Sibilia, M. Bertolotti, R. Ramponi, R. Osellame, M. Marangoni, J.W. Haus, M. Scalora, M.J. Bloemer, and C.M. Bowden, J. Opt. Soc. Am. B {\bf 19}, 2102 (2002).


\bibitem{OL2} G. Purvinis, P.S. Priambodo, M. Pomerantz, M. Zhou, T.A. Maldonado, and R. Magnusson, Opt. Lett. {\bf 29}, 1108 (2004).


\bibitem{PRL94}  B. Pettinger, X. Bao, I.C. Wilcock, M. Muhler, and G. Ertl, Phys. Rev. Lett.
{\bf 72}, 1561 (1994).

\bibitem{Chiang1} H.-P. Chiang, P.T. Leung, and W.S. Tse, J. Phys. Chem. B {\bf 104}, 2348 (2000).

\bibitem{Neeves} A.E. Neeves and M.H. Birnboim, J. Opt. Soc. Am. B
 {\bf 6}, 787 (1989).

\end{references}
\end{document}